\documentstyle[preprint,aps,psfig]{revtex}

\begin{document}
\draft
\preprint{\today}
\title
{Evaluation of the BCS Approximation for the Attractive Hubbard Model in One 
Dimension}
\def \bk{{\bf k}}
\def \bkp{{\bf k^\prime}}
\def \bQ  {{\bf Q}}
\def \bp {{\bf p}}
\def \bq {{\bf q}}
\def \br {{\bf r}}
\def \iwm {i\omega_m}
\def \iwmp {i\omega_{m^\prime}}
\def \kiwm {\bk,i\omega_m}
\def \kiwmp {\bk^\prime,i\omega_{m^\prime}}
\def\wtilde{\tilde {\omega }}
\def\YBCO{YBaCu_3O_{7-x}}
\def\BKBO{Ba_{1-x}K_xBiO_3}
\def\bea {\begin{eqnarray}}
\def\eea {\end{eqnarray}}
\def\be {\begin{equation}}
\def\ee {\end{equation}}
\def\eqnum#1{\eqno (#1)}
\def\sup{superconductivity }
\def\high{high $T_c$ superconductivity }
\author{F. Marsiglio}
\address
{Neutron \& Condensed Matter Science\\
AECL, Chalk River Laboratories, Chalk River, Ontario, Canada K0J 1J0,\\
Dept. of Physics \& Astronomy, McMaster University,
Hamilton, Ontario L8S 4M1}

\maketitle
\begin{abstract}
The ground state energy and energy gap to the first excited state are
calculated for the attractive Hubbard model in one dimension using both
the Bethe Ansatz equations and the variational BCS wavefunction. 
Comparisons are provided as a function of coupling strength and
electron density. While the ground state energies are always in
very good agreement, the BCS energy gap is sometimes incorrect by
an order of magnitude, particularly at half-filling. Finite size effects
are also briefly discussed for cases where an exact solution in the 
thermodynamic limit is not possible. In general, the BCS result for the
energy gap is poor compared to the exact result.
\end{abstract}
\bigskip
\noindent {\bf 1.  INTRODUCTION}
\par
The discovery of high-temperature superconductivity has motivated a
considerable effort over the last decade. In particular, many theoretical
models have been proposed to explain both the superconductivity and
some unusual normal state properties \cite{grenoble}. As far as
superconductivity is concerned, model building has proceeded on at least
two paths. The first includes ``realistic'' model Hamiltonians, often
two dimensional (since the CuO$_2$ planes have been deemed to be the 
essential structure for electron conduction). The price of such realism
is that exact solutions are impossible, and a mean field or BCS \cite{bcs}
solution is required. The second path attempts to remedy this deficiency by
seeking exact solutions to a restricted class of models, often in one (and
more recently in infinite) dimension. These models are often unrealistic,
and are sometimes exactly solvable only in restricted parameter regimes.
Attempts to bridge the gap between these two modes of theoretical
work are becoming more commonplace in recent years. Two examples
are spin fluctuation theories, where Fluctuation Exchange
calculations are often compared to Monte Carlo calculations for the
Hubbard model \cite{scalapino}, and Hubbard-like models that
include occupation-dependent hopping rates \cite{airoldi}. In this latter
example BCS calculations have been compared with exact diagonalization
studies on finite systems.\par

In these and other examples, comparison of BCS or some other approximation
scheme with exact results has always been hampered by some deficiency,
often finite size effects (one exception is mentioned below).
Yet an exact solution to the attractive Hubbard model
in one dimension has been available for
almost thirty years, via Bethe Ansatz techniques \cite{lieb}. \par

In this paper
a thorough investigation of the attractive Hubbard model will be presented,
using both Bethe Ansatz techniques and BCS solutions. The
study is necessarily confined to one dimension. It is not yet
clear to what extent conclusions obtained
in one dimension carry over to higher dimensions.
Nonetheless, we were encouraged by the results of Quick {\it et al.}
\cite{quick}, who performed a study similar to the present one
using a model of a one-dimensional electron-gas with pairwise-attractive
delta-function interactions, first solved exactly by Gaudin \cite{gaudin}
and Yang \cite{yang}. This is a one parameter model with a dimensionless
coupling constant inversely proportional to the electron density.
Quick {\it et al.} \cite{quick} found that BCS theory gave an accurate
estimate of the {\it ground state energy} over all ranges of the
coupling constant. Quantum fluctuations are most severe
in one dimension, so the agreement should improve as the dimension
is increased.\par

Part of the purpose of the present study is to test the 
accuracy of BCS as ``an interpolation scheme'' \cite{nozieres,leggett}
over all coupling strengths. As already noted in Ref. \cite{nozieres},
BCS theory is exact in the weak and strong coupling limits. The
present model has {\it two parameters}, electron density as well as
coupling strength, and the question of how well the interpolation scheme 
works as a function of electron density is addressed.
Moreover, the gap to the first 
excited state will also be calculated, as this quantity provides a much more
revealing test of the BCS approximation than does the ground state
energy. \par

We begin by outlining the model and the theoretical treatments in the
following section. For the exact solution we rely on previous work
by Lieb and Wu \cite{lieb} and by Bahder and Woynarovich \cite{bahder}.
We mention some trivial but important points regarding the solution of
the resulting equations, for both the ground state energy, and the
excitation gap. For completeness we also outline the BCS solution, which
is straightforward. Results are presented in the following section,
followed by a summary. \par
\bigskip
\noindent {\bf 2. FORMALISM}
\par
\medskip
{\bf A. The Model} \par
\medskip
The one dimensional Hubbard Hamiltonian is given by 
\be
H = -t \sum_{i,\sigma} \bigl( c^\dagger_{i\sigma}
c^{\phantom{dagger}}_{i+1,\sigma} + \, \, c^{\dagger}_{i+1,\sigma}
c^{\phantom{dagger}}_{i\sigma} \bigr)
+ \, \,
U\sum_i n_{i\uparrow}n_{i\downarrow},
\label{hubbham}
\ee
\noindent where the operator $c^{\dagger}_{i\sigma}$ ($c_{i\sigma}$)
creates (annihilates) an electron with spin $\sigma$ at site $i$ and
$n_{i\sigma}$ is the electron number operator with spin $\sigma$ at site
$i$. The parameter $t$ is the hopping rate for electrons and merely
sets the overall energy scale.
The parameter $U$ is negative in the attractive case, so that
$|U|/t$ is a dimensionless measure of the coupling strength. The second 
parameter is the electron density, $n \equiv {1 \over N}\sum_{i,\sigma} 
<n_{i\sigma}>$, where $N$ is the number of lattice sites, and the
expectation value is taken with respect to the ground state. \par

This model has particle-hole symmetry, so we will concern ourselves
with electron densities less than half-filling. In addition, symmetries
exist between the repulsive and attractive Hubbard models, such that
\cite{lieb}
\be
E(M,M^\prime;-|U|) = -M|U| \, + \, E(M,N-M^\prime;|U|)
\label{hubbsym}
\ee
\noindent where $E(M,M^\prime;|U|)$ is the ground state energy of $H$
(with $U = |U|$)
with $M$ down-spins and $M^\prime$ up-spins ($M + M^\prime = N_e$, the total
number of electrons). The total z-component of electron spin, $S_z$,
is given by $S_z \equiv {1 \over 2}(M^\prime - M)$. For the ground state,
$S_z = 0$.\par
\bigskip
{\bf B. The Lieb-Wu Equations: Finite Systems} \par
\medskip
An exact solution to the Hubbard model was first provided by Lieb and
Wu \cite{lieb}, using Bethe Ansatz techniques. A particularly enlightening 
derivation is provided by Sutherland in Ref. \cite{sutherland}. The
resulting equations to determine the wavevectors of the $N_e$ electrons
are:
\be
k_j = {2\pi \over N}I_j + {1 \over N} \sum_{\beta=1}^M \theta(2\sin{k_j}
-2\Lambda_\beta), \, \, \, j=1,2,3,\ldots,N_e.
\label{liebwu_a} \\
\ee
\noindent An auxiliary set of $M$ real numbers, the $\Lambda_\alpha$'s
are determined through the set of equations
\be
\sum_{j=1}^{N_e} \theta(2 \sin{k_j} - 2\Lambda_\alpha) = 2\pi J_\alpha
\, - \, \sum_{\beta=1}^M \theta(\Lambda_\alpha - \Lambda_\beta),
\, \, \, \alpha=1,2,3,\ldots,M,
\label{liebwu_b}
\ee
\noindent where
\be
\theta(p) \equiv -2\tan^{-1}(2pt/U), \, \, \, -\pi \le \theta < \pi.
\label{liebwu_def}
\ee
\noindent Here the $I_j$ are consecutive integers (half-odd integers)
if $M$ is even (odd), and the $J_\alpha$ are consecutive integers
(half-odd integers) if $M^\prime$ is odd (even). For the ground state we 
choose them to be (separately) clustered around zero (both negative
and positive). In general, for finite size systems, their sum
(${1 \over N} (\sum_j I_j + \sum_\alpha J_\alpha)$) does {\it not}
give zero, so that the total momentum of the ground state in some sectors
is not zero. For the attractive Hubbard model $M$ and $M^\prime$ are
kept close to one another so that the maximum number of electrons with
opposite spin can pair to take advantage of the attractive $|U|$. \par

These equations can be iterated to convergence using a standard
Newton-Raphson algorithm \cite{recipes}, although some damping of 
solutions and judicious starting points are sometimes required.
Once the $k_j$ have been obtained, the ground state energy is given
by
\be
E = -2t \sum_{j=1}^{N_e} \cos{k_j}.
\label{bethe_gse}
\ee
\noindent One way to determine the energy of the first excited state
is to compare the energy of two isolated systems
with $N_e$ electrons in one and $N_e + 2$ in the other, with two
isolated systems with $N_e + 1$ in each. This yields the pair binding
energy,
\be
\epsilon_b(N_e) = 2E(N_e + 1) - E(N_e) - E(N_e + 2).
\label{pair_energy}
\ee
\noindent Here we have used a single argument for the energy to indicate
the total number of electrons --- it is understood that $M$ and
$M^\prime$ are either equal (if $N_E$ is even) or differ by one (if
$N_e$ is odd). The single particle gap, $\Delta(N_e)$, is then given by
\be
\Delta(N_e) = \epsilon(N_e)/2.
\label{bethe_gap}
\ee
\noindent In this model the gap value is positive for $N_e$ even and negative
for $N_e$ odd, so that the overall curvature is in fact non-negative
(see the second reference in \cite{airoldi}). Thus, to calculate the gap
value in this manner requires a distinction between even and odd numbers
of electrons, a distinction which is unfortunately lost in the
thermodynamic limit. Nonetheless, finite size effects can be monitored
by increasing the system size until convergence is achieved. This
becomes quite time-consuming, however, for weak coupling (where large
system sizes are required for convergence), so that
an alternative method is required, as discussed next. \par
\bigskip
{\bf C. The Lieb-Wu Equations: Bulk Limit} \par
\medskip
As $N \rightarrow \infty$, $k$ and $\Lambda$ become distributed
throughout the first Brillouin zone and the real axis, respectively,
with density functions $\rho(k)$ and $\sigma(\Lambda)$, respectively
\cite{sutherland}. The above equations become the Lieb-Wu integral equations:
\be
\rho(k) = {1 \over 2\pi} + {\cos{k} \over \pi} \int_{-B}^B {|U|/4t \over
(U/4t)^2 + (\lambda - \sin{k})^2 }\sigma(\lambda) \, d\lambda,
\label{liebwuint_a}
\ee
\noindent with $\sigma(\lambda)$ determined self-consistently from
\be
{1 \over \pi} \int_{-Q}^Q {|U|/4t \over
(U/4t)^2 + (\lambda - \sin{k})^2 }\rho(k) \, dk = \sigma(\lambda) + 
{1 \over \pi} \int_{-B}^B { |U|/2t \over (U/2t)^2 +
(\lambda - \lambda^\prime)^2 } \sigma(\lambda^\prime) \, d\lambda^\prime,
\label{liebwuint_b}
\ee
\noindent where $Q$ and $B$ are determined by the following sum rules:
\be
\int_{-Q}^Q \rho(k) \, dk = 1 - 2s
\label{number_sum}
\ee
\be
\int_{-B}^B \sigma(\lambda) d\lambda = {n \over 2} - s
\label{spin_sum}
\ee
\noindent with $n \equiv N_e/N$ and $s \equiv S_z/N$. Finally the energy
per lattice site is given by
\be
{E \over N} = -|U|\bigl({n \over 2} - s \bigr) - 2t \int_{-Q}^Q \cos{k} \,
\,
\rho(k)\, dk
\label{liebwuint_energy}
\ee
\noindent Note that we have
already utilized Eq. (\ref{hubbsym}) in these equations; hence they apply to
an attractive Hubbard model with electron density $n$ and
magnetization $s$. The ground state is given by $s = 0$, so that
$Q = \pi$, i.e. the entire Brillouin zone is occupied. These equations
converge more easily than do their discrete counterparts given above, i.e.
simple iteration is sufficient, and Newton-Raphson (and therefore the 
inversion of a large matrix) is not required. \par

However, in principle, information concerning the gap is lost, as the
distinction between even and odd numbers of electrons is no longer possible
in the bulk limit. Nonetheless, Bahder and Woynarovich \cite{bahder}
use the trick of applying an external magnetic field. Minimization of
the energy then determines the magnetization as a function of applied
field. The magnetization is zero below a critical value of the field.
The energy associated with this critical field gives the gap in the 
spin excitation spectrum. It turns out that half this energy corresponds
exactly with the gap as defined by Eqs. (\ref{pair_energy}) and
(\ref{bethe_gap}). This is not a priori necessary, as the spin-gap is
defined by flipping a spin with fixed electron number, whereas the previous
gap was defined by changing the number of electrons in the system. \par

To determine the gap in the bulk limit, we use a different
procedure, which initially follows the method outlined in Ref. \cite{bahder}
(see their section IV where they carry out an analytical calculation in
the low and high density limits). As already mentioned,
the spin-gap can be defined as the energy 
required to flip one spin. Therefore we are interested in the energy 
difference between the $s = 0$ state and the $s = 1/N$ state where $N$
is taken to approach $\infty$. Since $Q = \pi$ for $s = 0$, we have from
Eq. (\ref{number_sum}) that $Q = \pi - 1/[N\rho(\pi)]$, using the even
symmetry of $\rho(k)$. Substitution into Eq. (\ref{liebwuint_b}) yields one
term on the left-hand side which is $O(1/N)$. Furthermore, $\rho(k)$
can now be eliminated by substituting Eq. (\ref{liebwuint_a}) into
the remaining integral from $-\pi$ to $\pi$ \cite{bahder}:
\be
{1 \over 2\pi^2}\int_{-\pi}^{\pi} {|U|/4t \over (U/4t)^2 + (\lambda -
\sin{k})^2 } - {2 \over \pi N} {|U|/4t \over (U/4t)^2 + \lambda^2} = 
\sigma(\lambda) + {1 \over \pi} \int_{-B}^B d\lambda^\prime
\sigma(\lambda^\prime) {|U|/2t \over (U/2t)^2 + (\lambda - \lambda^\prime)^2}.
\label{eqn1}
\ee
\noindent To first order in $1/N$ there is no change in the density
function $\rho(k)$; however, the same is not true for $\sigma(\lambda)$.
To see this we substitute $s= 1/N$
into Eq. (\ref{spin_sum}), and allow $B$ to adjust ($B \rightarrow
B_\circ + \delta B$) to the flipped spin (here the subscript $\circ$ signifies
the solution for $s = 0$). The important point is that the electron density 
$n$ remains constant. Then
\be
\delta B = {1 \over 2\sigma(B_\circ)} \bigl\{ \int_{-B_\circ}^{B_\circ}
\sigma(\lambda) \, d\lambda - {n \over 2} + {1 \over N} \bigr\}
\label{delb}
\ee
\noindent where again we have utilized the even symmetry of $\sigma(\lambda)$.
The two first terms within the braces {\it do not cancel} because
there are $O(1/N)$ corrections to the density function $\sigma(\lambda)$.
This is most clearly seen at half-filling, where $\delta B = 0$ and
the sum rule has apparently changed by $O(1/N)$. Away from half-filling
$\delta B$ is $O(1/N)$. Substitution of Eq. (\ref{delb})
into Eq. (\ref{eqn1}) results in 
\bea
\sigma(\lambda) & = &
{1 \over 2\pi^2}\int_{-\pi}^{\pi} dk {|U|/4t \over (U/4t)^2 + (\lambda -
\sin{k})^2 }
- {1 \over \pi} \int_{-B_\circ}^{B_\circ} d\lambda^\prime
\sigma(\lambda^\prime) {|U|/2t \over (U/2t)^2 + (\lambda - \lambda^\prime)^2}
\nonumber \\
& &
- {2 \over \pi N} {|U|/4t \over (U/4t)^2 + \lambda^2}
+ {\delta B \over \pi} \sigma(B_\circ)
\biggl\{
{|U|/2t \over (U/2t)^2 + (\lambda - B_\circ)^2}  +
{|U|/2t \over (U/2t)^2 + (\lambda + B_\circ)^2}
\biggr\}
\label{eqn2}
\eea
\noindent where every term except the first on the right-hand-side
contains (explicit or implicit) terms with $O(1/N)$.
This equation is iterated to convergence (with high precision)
with some choice for $N$ (eg. $N = 10^4$ or $10^5$). The equation for the 
energy, Eq. (\ref{liebwuint_energy}) becomes
\bea
{E \over N} & = & -|U|{n \over 2} - {2t \over \pi} \int_{-\pi}^{\pi}
dk \, \cos^2{k} \int_{-B_\circ}^{B_\circ} d\lambda \, \sigma(\lambda)
{|U|/4t \over (U/4t)^2 + (\lambda - \sin{k})^2 }
\nonumber \\
&  &
+ {|U| - 4t \over N} - \delta B \, \sigma(B_\circ)
\biggl\{
{|U|/4 \over (U/4t)^2 + (B_\circ - \sin{k})^2}  +
{|U|/4 \over (U/4t)^2 + (B_\circ + \sin{k})^2}
\biggr\}.
\label{eqn3}
\eea
\noindent The gap energy is then identified as the coefficient of
the $1/N$ term in the energy,
\be
{E \over N} \equiv {E_\circ \over N} + { 2\Delta \over N}.
\label{bethe_gap2}
\ee
\noindent The first term on the right-hand-side is obtained by solving
the integral equations with $N \rightarrow \infty$, i.e. Eqs.
(\ref{liebwuint_a}-\ref{liebwuint_energy}). Then the left-hand-side is 
obtained by solving Eqs. (\ref{delb}-\ref{eqn3}) for $N = 10^4, 10^5$,
etc., until the gap extracted from Eq. (\ref{bethe_gap2}) has converged. \par
\bigskip
{\bf D. The BCS Equations} \par
\medskip
The derivation of the BCS equations is given in many places
\cite{schrieffer} and will not be presented here. They are:
\be
1 = {|U| \over N} \sum_k {1 \over 2E_k}
\label{bcs1}
\ee
\noindent with
\be
E_k = \sqrt{(\epsilon_k - \mu - |U|n/2)^2 + \Delta_{\rm BCS}^2}
\label{bcs2}
\ee
\noindent and $\epsilon_k = -2t\cos{k}$, where the sums are carried out
over the first Brillouin zone ($-\pi < k\le \pi$). We use a finite lattice
for simplicity, large enough for the gap to have converged to its bulk 
limit. An auxiliary equation to determine the electron number density
is also required,
\be
n = 1 - {1 \over N} \sum_k {\epsilon_k - \mu - |U|n/2 \over E_k}.
\label{bcs3}
\ee
\noindent For a given $|U|$ and $n$ these equations are iterated to 
convergence to determine $\mu$ and $\Delta_{\rm BCS}$.
If the chemical potential lies within the band, then the gap to the
first excited state, $\Delta_\circ$,  is given by
$\Delta_{\rm BCS}$. Otherwise, the gap is defined by the quasiparticle
energy at the bottom of the band, i.e.
\bea
\Delta_{\circ} & = & \Delta_{\rm BCS} \phantom{aaaaaaaaaaaaaaaaaaaaaaaa}
{\rm for} \ \mu + |U|n/2 > -2t
\nonumber \\
& = & \sqrt{(2t + \mu + |U|n/2)^2 + \Delta_{\rm BCS}^2}
\phantom{aaaaa} {\rm for} \ \mu + |U|n/2 < - 2t
\label{mingap}
\eea
\noindent $\Delta_\circ$  is to be directly compared to the gap discussed
in the previous sections. Finally, the total energy
is given by
\be
{E_{\rm BCS} \over N} = {1 \over N} \sum_k \epsilon_k \biggl(
1 - {\epsilon_k - \mu - |U|n/2 \over E_k} \biggr) - |U| \biggl({n \over 2}
\biggr)^2 - { \Delta_{\rm BCS}^2 \over |U|}.
\label{bcs4}
\ee
\noindent Note that the Hartree term is not required in the BCS
gap equations (and is often omitted),
but it provides an important contribution to the total energy. \par
\bigskip
\noindent {\bf 3. RESULTS}
\par
\bigskip
{\bf A. The Ground State Energy} \par
\medskip
In the dilute limit the BCS solution of the gap equations
is exact for all coupling strengths
\cite{nozieres}. Of course the BCS approximation
also reduces to the non-interacting
result in the limit of weak coupling, for any electron density. Finally,
BCS theory also becomes exact in the limit of strong coupling, as
was appreciated for both the electron gas \cite{eagles} and the lattice gas
\cite{nozieres}.\par

In Fig. 1 the ground state energy is shown as a function of electron 
density for various values of $U$. We have shown some results for
repulsive $U$ \cite{shiba} along with the analytical results for
$U = 0$ ($E/N = -{4 \over \pi} \sin{n \pi \over 2}$) and
$U = \infty$ ($E/N = -{2 \over \pi} \sin{n \pi}$). Note that agreement
is best in weak coupling and in strong coupling, particularly for
low electron density. Nonetheless, the BCS energy follows the exact result
much more closely in the intermediate regime for this model than appears
to be the case for the electron gas model \cite{quick}. In fact a
strong coupling expansion shows that the ground state energy has
corrections of order $(t/U)^2$ in both the exact and BCS solutions,
whereas for the gas model the BCS solution contained a term linear
in the inverse coupling constant which was not present in the exact
solution \cite{quick}.\par

In Fig. 2 we show the ground state energy as a function of coupling 
strength, $|U|$ for half-filling ($n = 1$) and quarter-filling ($n = 0.5$).
Here it is clear that the BCS theory is most inaccurate for intermediate
coupling strength, i.e. $|U| \approx {\rm bandwidth}$. Nonetheless the 
maximum deviation is rather small, $\approx 4 \% $, which occurs at
half-filling near $|U| = 4t$. Thus, on the basis of ground state energy
calculations, BCS theory appears to be a very accurate theory, even
in one dimension. However, in the next subsection we show that the BCS result
for the energy gap is far less accurate. \par
\bigskip
{\bf B. The Energy Gap} \par
\medskip
The calculation of the energy gap is a much more sensitive test of
the accuracy of BCS theory. In Fig. 3 we show the gap as a function
of electron density for various values of the interaction strength, $|U|$.
As already stated, the BCS gap is exact in the low density limit, regardless
of the coupling strength. However, as the electron density increases, the
true gap {\it decreases} for all coupling strengths, while the BCS gap
($\Delta_{\circ}$) from Eq. (\ref{mingap}) {\it increases} for all
coupling strengths. The true gap decreases monotonically as a  function
of electron density, whereas the BCS gap is in general non-monotonic,
exhibiting a maximum at some intermediate electron density, which
is a function of coupling strength. For sufficiently strong coupling,
the BCS gap is monotonically increasing, so that the maximum occurs
at half-filling. The most serious errors occur near half-filling, 
particularly in the weak coupling limit, where BCS theory vastly
overestimates the value of the gap. \par

At half-filling certain analytical
results can be obtained. For the true gap we obtain
\be
\Delta(n=1) = {|U| \over 2} - 2t + 4t\int_0^\infty {d\omega \over \omega}
{J_1(\omega) \over 1+ e^{\omega|U|/2t}}
\label{gap_half}
\ee
\noindent where $J_1(\omega)$ is the Bessel function of the first
kind of order 1. This result can be written in a more useful form
\cite{leewhiting},
\be
\Delta(n=1) = {8t \over \pi} \sum_{m=1}^\infty {1 \over 2m-1}
K_1\biggl({2\pi t\over |U|}(2m-1) \biggr),
\label{graham}
\ee
\noindent where $K_1(x)$ is the first order modified Bessel function.
In weak coupling this reduces to an exponential contribution:
\be
\Delta_{\rm weak} = {4 \over \pi}\sqrt{|U|t} \exp{\biggl\{ {-2\pi t \over
|U|} \biggr\}}.
\label{gap_half_weak}
\ee
\noindent In strong coupling the first two terms of Eq. (\ref{gap_half})
dominate.\par

The BCS integral can also be performed analytically: we obtain, at
half-filling,
\be
{2\pi t \over |U|} = {1 \over \sqrt{1 + \delta^2}} K\biggl({1 \over
1 + \delta^2}\biggr),
\label{bcsgap_half}
\ee
\noindent where $\delta \equiv {\Delta_\circ \over 4t}$ and
$K(k) \equiv \int_0^{\pi/2} {1 \over \sqrt{1 - k\sin^2{\theta}} }d\theta$
is the complete elliptic integral of the first kind. Note that
$\Delta_\circ = \Delta_{\rm BCS}$ at half-filling. In weak coupling we
obtain
\be
\Delta_{\circ(\rm weak)}  = 8t\exp\biggl( - {2\pi t \over |U|}\biggr)
\label{bcsgap_half_weak}
\ee
\noindent while in strong coupling we have
\be
\Delta_{\circ(\rm strong)}  = {|U| \over 2} - {2t^2 \over |U|}.
\label{bcsgap_half_strong}
\ee
\noindent Thus, the weak coupling BCS gap at half-filling
has the familiar activated form,
albeit with incorrect prefactors. In strong coupling the first
order correction (of order unity) is absent in the BCS solution.\par

Finally, at zero filling, the exact solution is known analytically
\cite{hirsch}:
\be
\Delta = -2t + \sqrt{\bigl( {|U| \over 2}\bigr)^2 + \bigl( 2t \bigr)^2},
\label{gap_zero}
\ee
\noindent and the BCS result is identical. Note that in weak coupling
the dependence on $|U|$ is quadratic. \par

In Fig. 4 we show the gap as a function of coupling strength for
zero-, quarter-, and half-filling. As already stated, for zero filling
BCS is exact, while for any non-zero filling, the deviations are as shown.
BCS theory overestimates the gap by a considerable degree over the
entire coupling range shown. Note that the variation of the exact gap with
electron density is not significant beyond $|U|/t \approx 2$, and, as 
remarked earlier, monotonically decreases with increasing electron density.
The dependence of the BCS gap on the electron density  is not monotonic,
and beyond $|U|/t \approx 3$ the ordering is incorrect. In Fig. 4b we
show the weak coupling regime in greater detail. In this regime the 
ordering with electron density is correct, but BCS theory greatly
overestimates the gap value, as shown.\par
\bigskip
{\bf C. Finite Size Effects} \par
\medskip
The attractive Hubbard model can be solved in the thermodynamic limit, and
therefore, for this study, finite size effects do not concern us.
However, most models require some sort of numerical solution, often
on small system sizes, and are therefore subject to finite size
effects. It is therefore of interest to examine how these
effects influence the results in both the exact and BCS solution
in the present model. To some extent the following conclusions can be
applied to other models where an exact solution in the bulk limit is not
possible.\par

To obtain the exact gap it is necessary to solve
Eqs.(\ref{liebwu_a}-\ref{bethe_gap}) for some fixed lattice size, $N$.
The BCS gap is obtained from Eqs.(\ref{bcs1}-\ref{mingap}), again for
a fixed lattice size, $N$. However, these equations utilize a grand
canonical ensemble, so that particle number is fixed only in an
average sense, through the chemical potential. Thus, all electron densities
are possible (except for certain densities where discontinuities may arise);
this raises the issue of how applicable the use of the grand canonical
ensemble is for small clusters. We partially address this issue here
by using the canonical ensemble for clusters with one, two, three and four
electrons. This has already been done for a tetrahedral cluster
\cite{falicov}, where, for symmetry reasons, the BCS  solution is exact
for all pair-fillings on a four-site cluster. We use the BCS wavefunctions
\bea
|\Psi_2> & = & \sum_k g_k c^\dagger_{k\uparrow} c^\dagger_{-k\downarrow}
|0>
\nonumber \\
|\Psi_4> & = & \sum_{k_1 \atop k_2 > k_1} C(k_1,k_2)
c^\dagger_{k_1\uparrow} c^\dagger_{-k_1\downarrow}
c^\dagger_{k_2\uparrow} c^\dagger_{-k_2\downarrow} |0>
\label{bcsproj}
\eea
\noindent for two and four electrons, respectively. On a chain of four
sites these would correspond to quarter- and half-filling. To accomodate
odd numbers of electrons we simply add another electron, whose momentum
then determines the total momentum of the ground state wavefunction. For
example, for three electrons,
\be
|\Psi_3>_q  =  \sum_{k \not= q} g^\prime_k
c^\dagger_{k\uparrow} c^\dagger_{-k\downarrow}
c^\dagger_{q\uparrow} |0>
\label{bcsproj3}
\ee
\noindent and the third electron has the important role of blocking
the pair from adopting the $(q,-q)$ pair state. Minimization with
respect to the relevant parameters yields equations for the $g_k$ or
$ C(k_1,k_2)$ from
which the minimum energy can be obtained. The gap energy can then be
obtained following the method outlined in Section 2.B. In contrast to
the grand canonical method for finite systems, this method generates
gap energies for values of the electron density that are commensurate with
the lattice size. \par

In Fig. 5 results for the gap are illustrated as a function of density
for $|U| = -2t$, using all the various methods discussed thus far. The open
symbols  represent exact solutions for various lattice sizes as indicated
in the figure caption. The gap generally decreases as the lattice
approaches the bulk limit. The same trend is observed in the BCS
solutions, indicated by the various curves. Finally, for $N = 4$,
the filled circle indicates the result for quarter filling
from the canonical BCS equations.\par

We have found quite generally that the use of the canonical ensemble
always improves the ground state energy, compared to that obtained
from the grand canonical ensemble. This is particularly true for
electron densities that correspond to an odd number of electrons, and
for low electron densities. In fact the conventional BCS theory
is variational with respect to $H - \mu N$, and not the Hamiltonian
alone, so that for small system sizes this theory gives
{\it lower} energies than the true ground state energies at low
electron densities. This does not violate the variational principle since
the BCS wavefunction used spans all particle numbers, and BCS theory
never gives a lower energy than the true ground state energy {\it for
any particle number}. In any event, this distasteful feature is
remedied by the use of the canonical ensemble. Of course, as the bulk
limit is approached these discrepancies disappear.
\par

However, the use of the canonical ensemble {\it does not} universally
improve the BCS gap energy for small systems. The case shown in Fig. 5
shows some improvement. For weaker coupling the improvement is
insignificant, while for stronger coupling the result from the
canonical solution is poorer than the grand canonical one.\par

It should be clear that finite size effects are significant only
in the intermediate to weak coupling regime (the two site system
reproduces the correct bulk limit in the strong coupling limit).
For the intermediate coupling case shown, the finite size results
could {\it mislead} one to believe that the gap energy {\it increases}
as electron density increases from zero. This behaviour is displayed
by the BCS results as well, and remains a feature of the BCS solution
in the bulk limit (solid line). However, the true energy gap is a 
monotonically decreasing function of electron density. Furthermore,
the gap always decreases as the bulk limit is approached, so that the
relative error obtained from a finite system study will always underestimate
the true error in the bulk limit. The error in the BCS gap at half-filling,
for example, is $20\%$ for a four site chain, while, in the bulk limit
it is $300\%$. This underestimate worsens for weaker coupling.\par

Finally, we should note that the finite size effects on the ground state
energy are not nearly as severe as just discussed for the energy gap.
For example, the ground state energy at half-filling for $N=8$ differs from
the bulk limit by about $1\%$ while the gap value is almost a factor
of three too large.\par
\bigskip
\noindent {\bf 4.  SUMMARY}
\par
We have carried out an evaluation of BCS theory for the attractive Hubbard
model in one dimension. This is certainly the simplest Hamiltonian
with many-body interactions for which an exact solution exists, so
that an evaluation in the bulk limit is possible. We hope that the
results of this paper can be used for more complicated Hamiltonians
where exact solutions are not possible.\par

The ground state energy is very accurately reproduced by the BCS
wavefunction, particularly at low electron densities, and for all coupling
strengths. In fact the ground state energy is least accurate at
intermediate coupling strengths, at half-filling, and here the error
is only a few per cent. This result is encouraging, and ought to
improve with increasing dimensionality.\par

The energy gap is not so accurately reproduced within BCS theory. There
is exact agreement in the dilute limit; however the exact result
monotonically decreases with electron density while the BCS
result increases, at least initially. There is also agreement in the
strong coupling limit, and BCS theory displays the same activated form
in the weak coupling limit, but with an incorrect prefactor. At
intermediate coupling strength the BCS gap overestimates the true
gap typically by a factor of two or more. The relative error decreases
(increases) with stronger (weaker) coupling. Over most of the coupling
strength regime the maximum error occurs at half-filling.\par

Finally, we have taken advantage of this exactly solvable case to
examine finite size effects. Insofar as these effects occur in other
more complicated models, they can be used as cautionary guidelines
in future finite system studies.\par

The relevance to higher dimensionality remains somewhat of an open
question. While the accuracy of the BCS ground state energy is most
encouraging, it is also clear from this study that this agreement is
not a good indicator of the accuracy of the BCS energy gap. On the other
hand the poor accuracy of the BCS gap in one dimension may well be
rectified by proceeding to higher dimension. It is hoped that future 
investigations of this issue can utilize this study as a benchmark.\par

\noindent {\bf Acknowledgements}\par
\medskip
\noindent I thank Takashi Nakatsukasa for helpful discussions and
Graham Lee-Whiting for his derivation of Eq. (\ref{graham}). Partial
support from the Natural Science and Engineering Research Council
(NSERC) of Canada is acknowledged.\par
\vfil\eject

\vfil\eject
\begin{figure}
\psfig{file=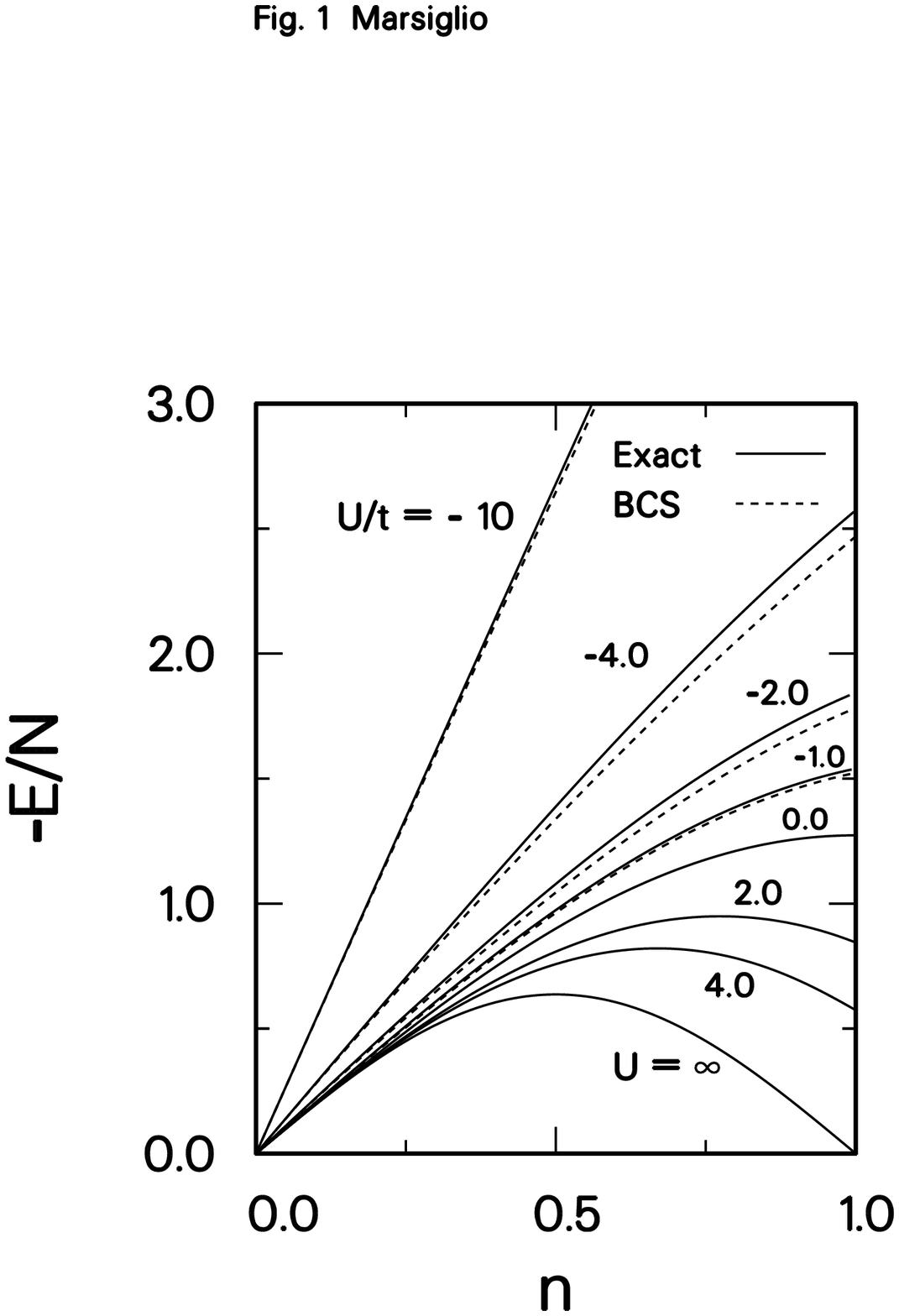,height=7.5in,width=7.in}
\vskip+0mm
\caption{The ground state energy as a function of electron density
for various values of the coupling strength, both repulsive ($U > 0$)
and attractive ($U < 0 $). BCS results are also shown for comparison.}

\end{figure}
\vfil\eject
\begin{figure}
\psfig{file=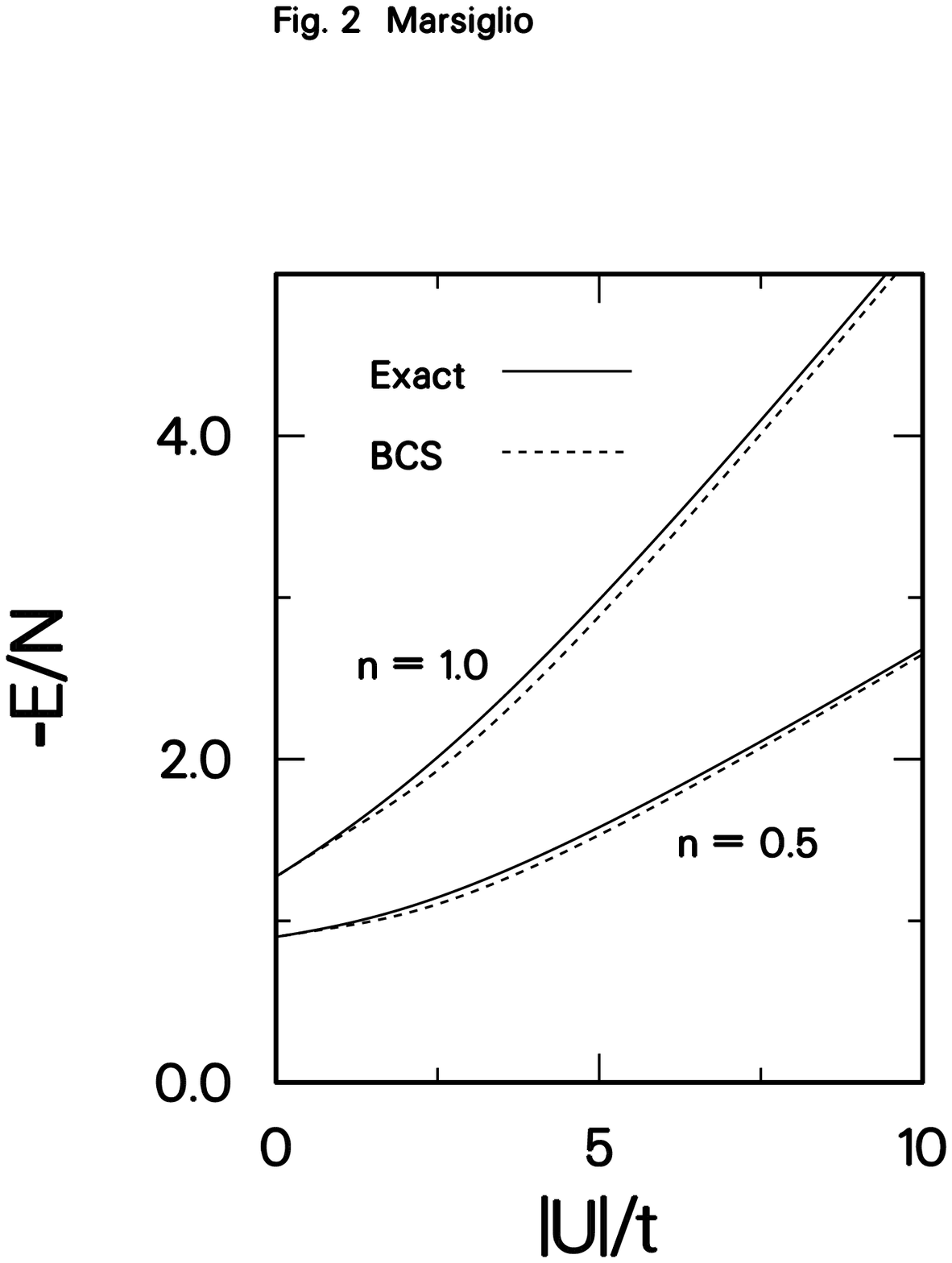,height=7.5in,width=7.in}
\vskip+0mm
\caption{The ground state energy as a function of the coupling
strength for quarter ($n = 0.5$) and half ($n=1.0$) filling, along
with the BCS results.}

\end{figure}
\vfil\eject

\begin{figure}
\psfig{file=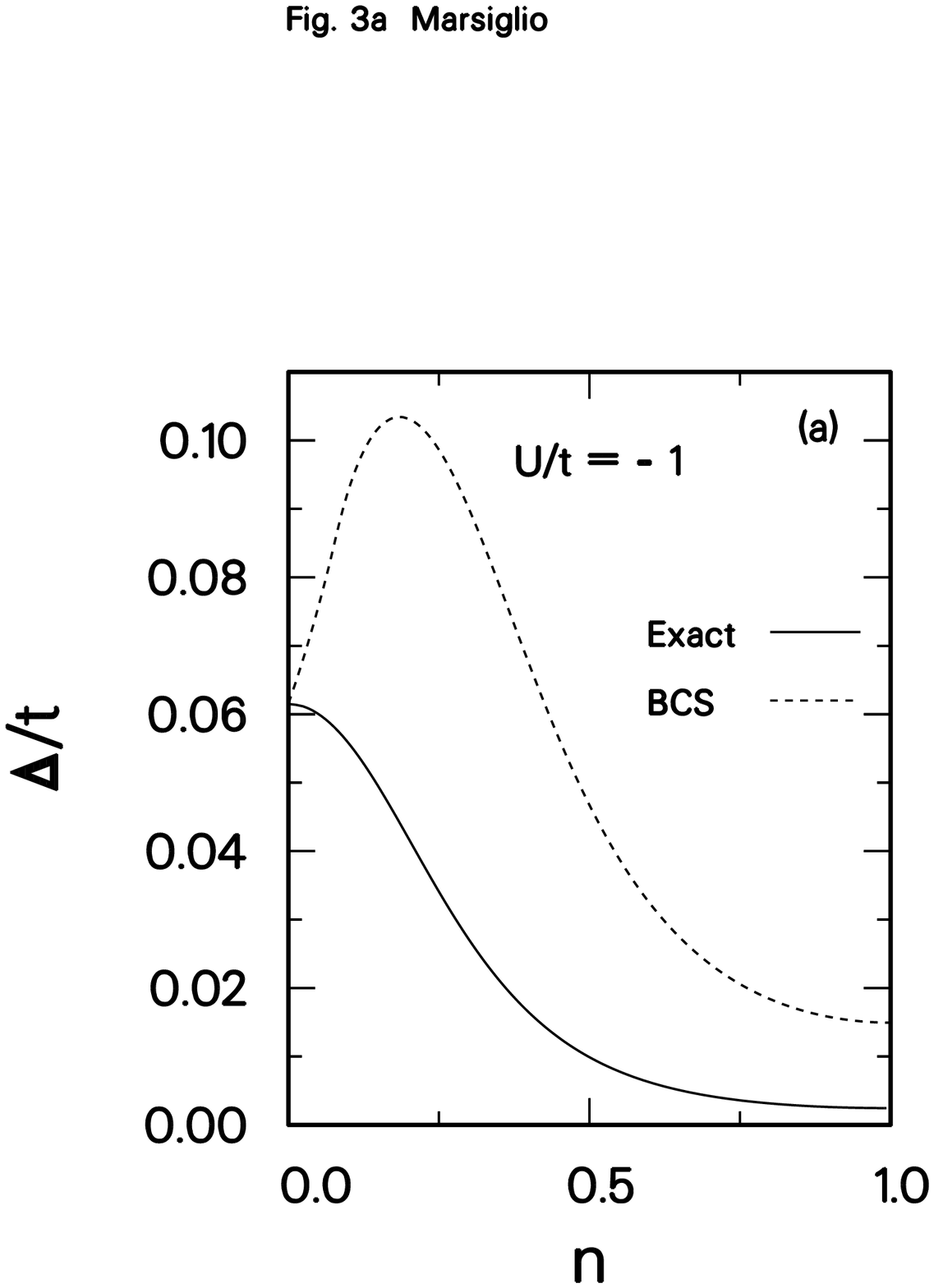,height=7.5in,width=7.in}
\vskip+0mm
\caption{The energy gap, $\Delta/t$ vs. electron density for
(a) $|U|/t = 1$, (b) $|U|/t = 2$, and (c) $|U|/t = 10$. The BCS result
(i.e. $\Delta_\circ/t$) is shown for comparison. }

\end{figure}
\vfil\eject
\begin{figure}
\psfig{file=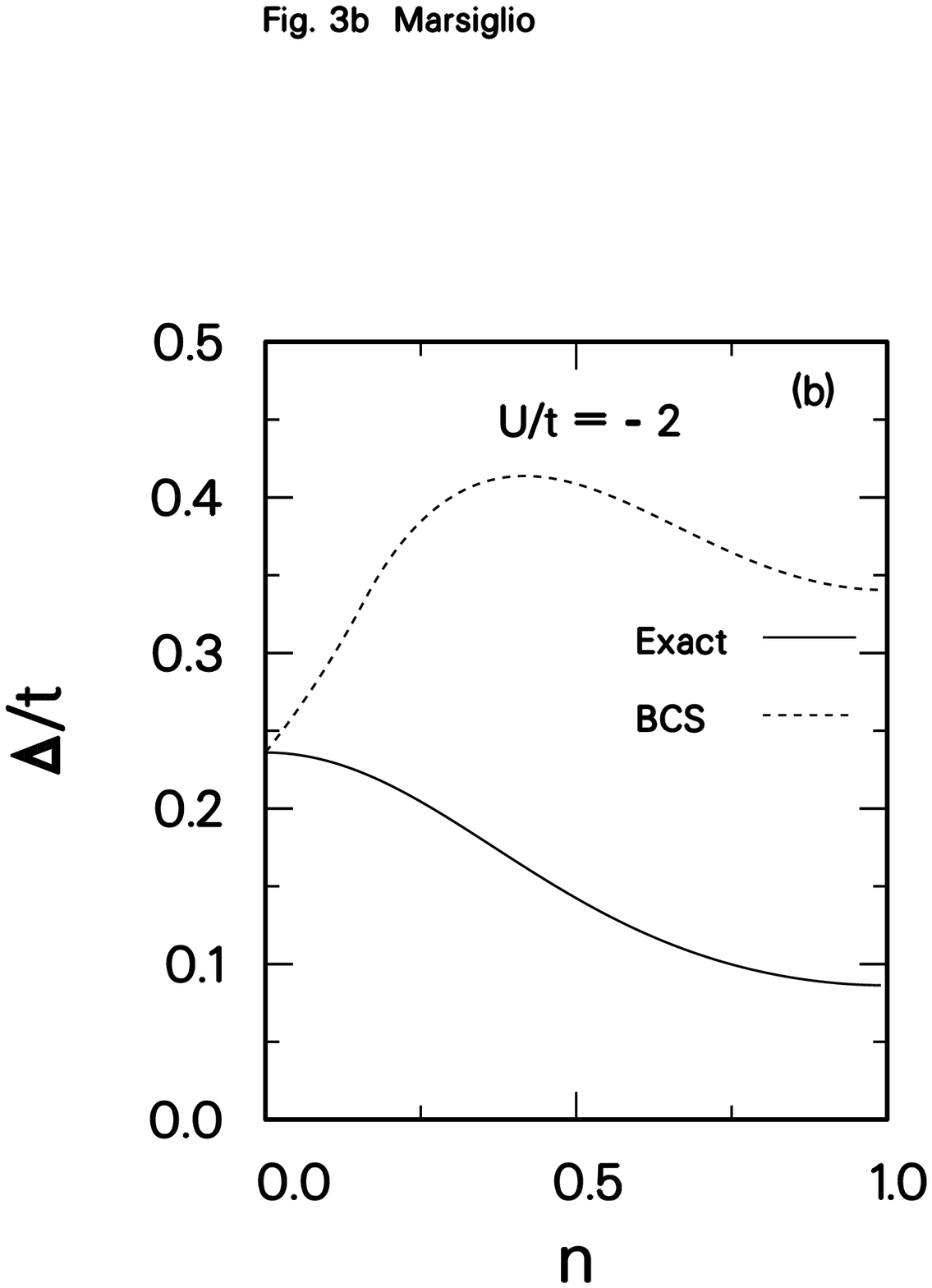,height=7.5in,width=7.in}
\vskip+0mm
%

\end{figure}
\vfil\eject
\begin{figure}
\psfig{file=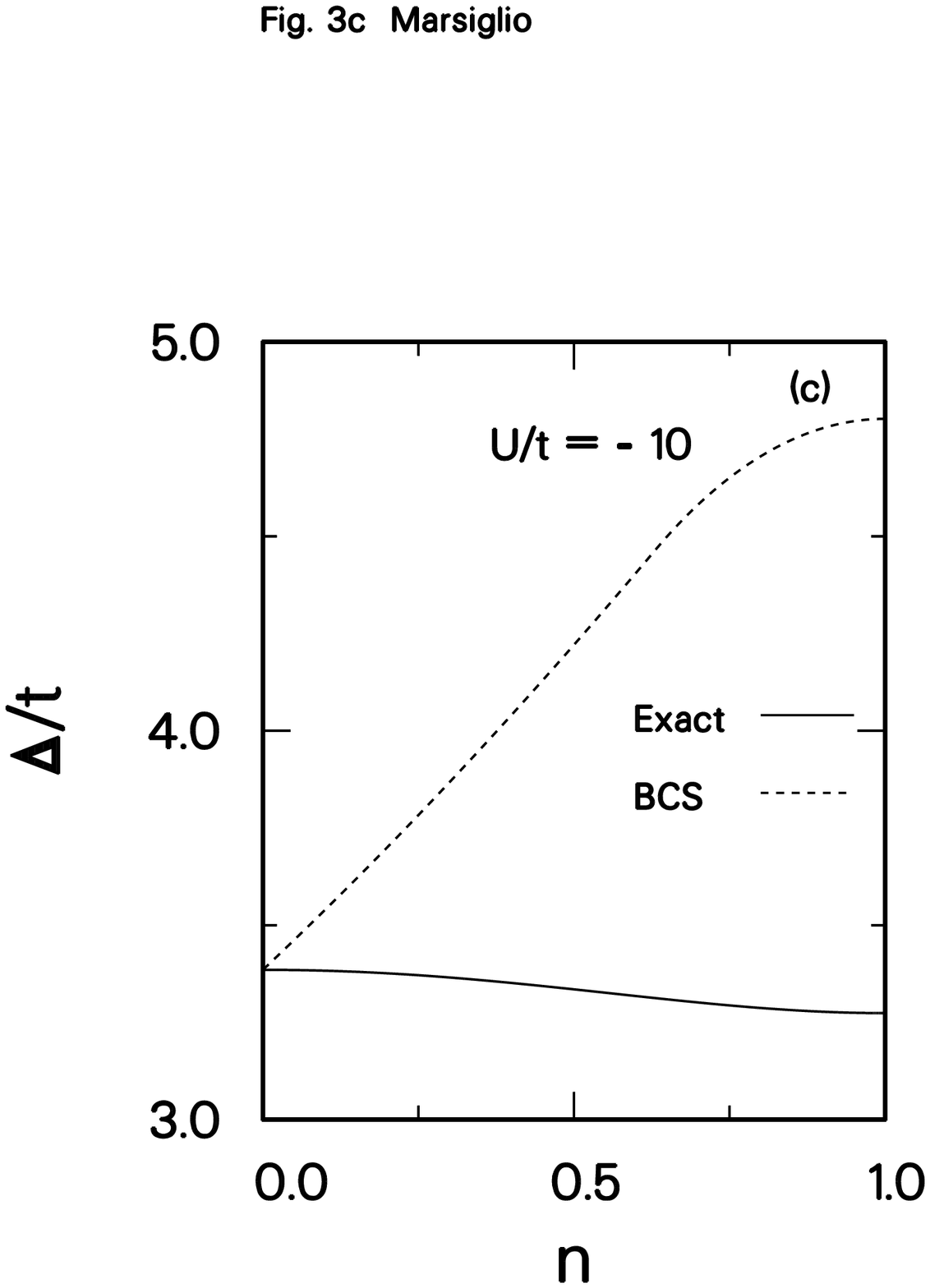,height=7.5in,width=7.in}
\vskip+0mm
%

\end{figure}
\vfil\eject
\begin{figure}
\psfig{file=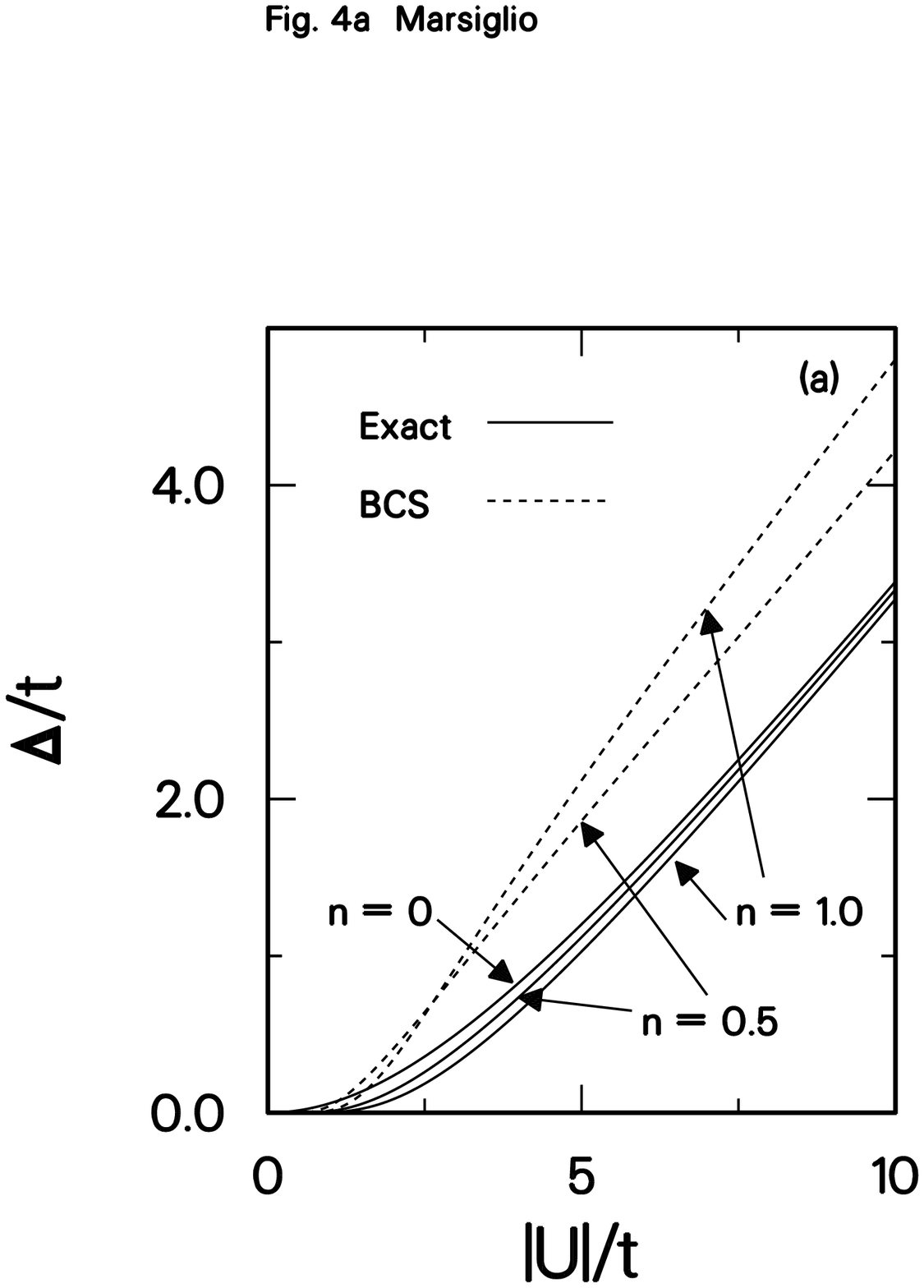,height=7.5in,width=7.in}
\vskip+0mm
\caption{(a) The energy gap, $\Delta/t$ vs. coupling strength, $|U|/t$,
for various electron densities. In (b) we provide an expanded version
of the weak coupling regime.}

\end{figure}
\vfil\eject
\begin{figure}
\psfig{file=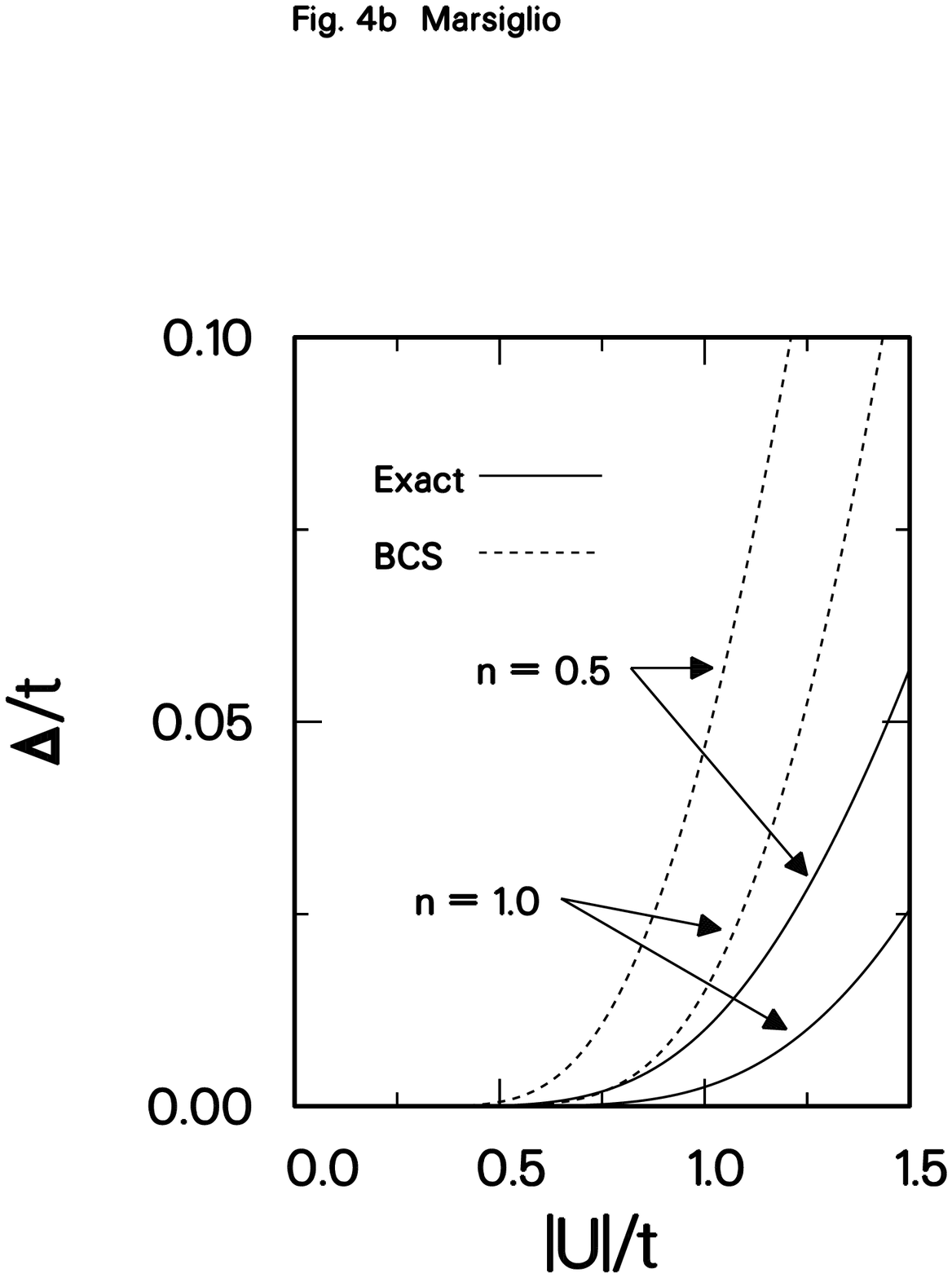,height=7.5in,width=7.in}
\vskip+0mm
%

\end{figure}
\vfil\eject
\begin{figure}
\psfig{file=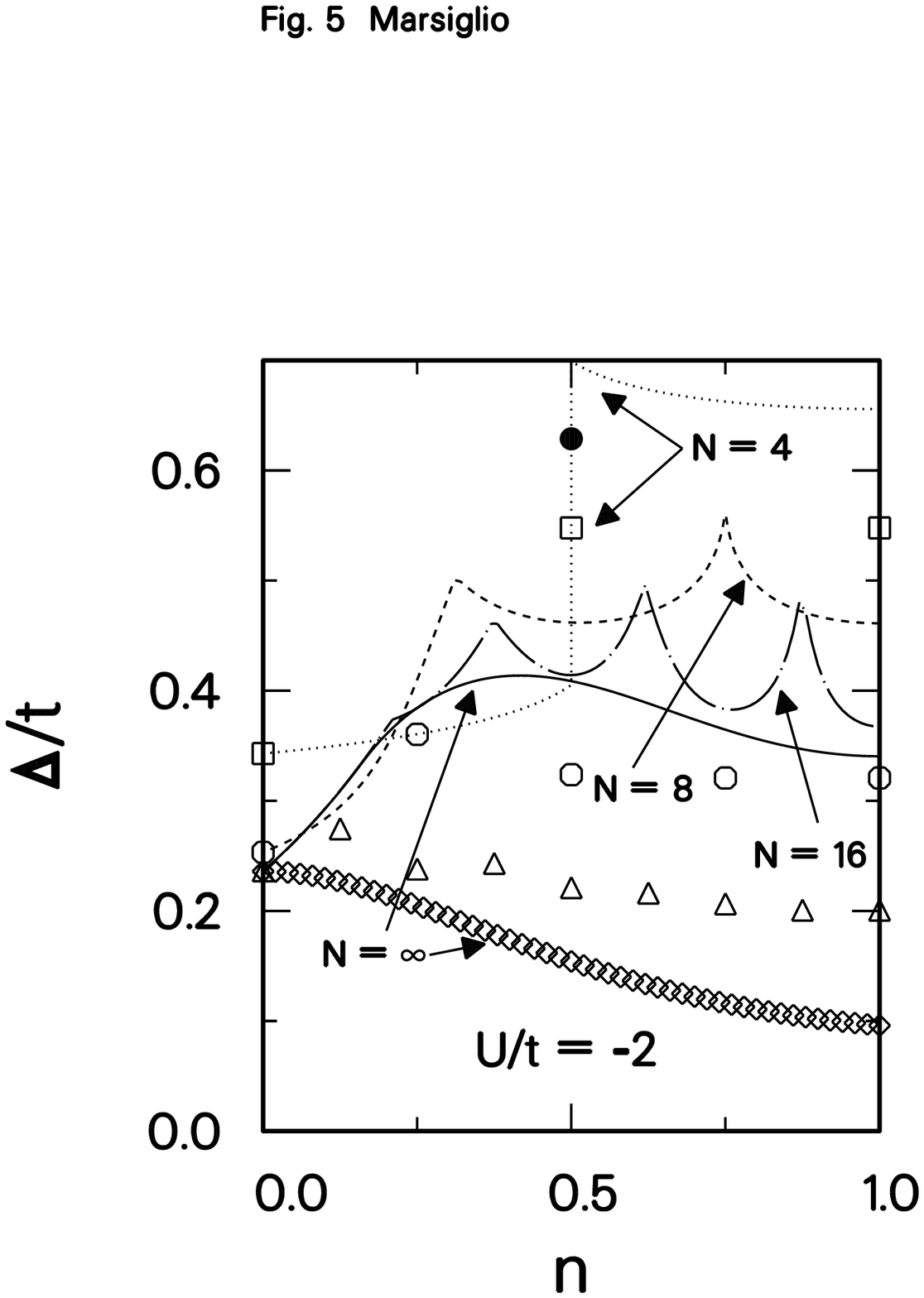,height=7.5in,width=7.in}
\vskip+0mm
\caption{The energy gap, $\Delta/t$ vs. electron density for
$|U|/t = 2$, for various chain sizes. The open symbols are from exact
results, with lattice size as indicated. The curves are BCS results
(using the grand canonical ensemble) with lattice sizes as indicated.
We
have also indicated the result for quarter-filling for $N = 4$
(filled circle), showing an improvement over the grand canonical
result.}

\end{figure}
\end{document}